\documentclass{article}
\usepackage{amsmath}
\usepackage{amssymb}
\usepackage{amsthm}
\usepackage{tikz}
\usepackage{caption}
\usepackage{subcaption}
\def\R{{\mathbb R}}

\begin{document}

\title{Phenomenology of an in-host model of hepatitis C}

\author{Alexis Nangue\\
Higher Teachers' Training College, University of Maroua\\
P.O. Box 55, Maroua, Cameroon\\  
and\\
Alan D. Rendall\\
Institut f\"ur Mathematik\\
Johannes Gutenberg-Universit\"at, Staudingerweg 9\\
D-55099 Mainz, Germany}

\date{}

\maketitle

\begin{abstract}
This paper carries out an analysis of the global properties of solutions of
an in-host model of hepatitis C for general values of its parameters. A
previously unknown stable steady state on the boundary of the positive orthant
is exhibited. It is proved that the model exhibits Hopf bifurcations and hence
periodic solutions. A general parametrization of positive steady states is
given and it is determined when the number of steady states is odd or even,
according to the value of a certain basic reproductive ratio. This implies,
in particular, that when this reproductive ratio is greater than one there
always exists at least one positive steady state. A positive steady state
which bifurcates from an infection-free state when the reproductive ratio
passes through one is always stable, i.e. no backward bifurcation occurs in
this model.
\end{abstract}

\section{Introduction}\label{intro}

Hepatitis C is a major health problem on a global scale. This is illustrated
by the following facts \cite{who}. There are 1.5 million infections per year.
Of those infected 30\% recover after a few months while the disease becomes
chronic in the remaining 70\% of cases. The estimated number of people with
chronic hepatitis C is 58 million. Chronic infection leads in many cases to
cirrhosis and liver cancer. In 2019 the number of people who died from
hepatitis C (mainly the long-term complications) was 290000. These numbers
may be compared with those for hepatitis B, a disease due to an unrelated virus
which also
affects the liver. In that case there are also 1.5 million infections per year.
The disease can become chronic, mainly in the case of childhood infections.
The estimated number of people with chronic hepatitis B is 296 million. As in
the case of hepatitis C  chronic infection leads in many cases to cirrhosis
and liver cancer. In 2019 the number of people who died from hepatitis B
(mainly the long-term complications) was 820000. These numbers may be compared
with those for a viral infection better known to the general public, HIV/AIDS.
In that case the estimated number of infections per year is again 1.5 million.
The disease always becomes chronic. The estimated number of people infected
with HIV is 37.7 million. In 2020 the number of people who died from the effects
of HIV was 680000.

There is no effective vaccine against hepatitis C. However there are treatments
with direct-acting antivirals (DAA) which can cure most chronic cases. They do
not suffer from the very long treatment times and severe side effects of earlier
treatments such as interferon $\alpha$ and ribavirin. The availability of the
new
treatments has been limited by their high costs. In the case of hepatitis B
there is an effective vaccine but there is no curative treatment. There are
drugs which may help to reduce viral load and hence long-term damage to the
liver. In the case of HIV there is neither an effective vaccine nor a curative
treatment. There are drugs which can be used to maintain the virus load on a
low level on a long-term basis (highly active anti-retroviral therapy, HAART).
There is a great need to understand these diseases better and in doing so it
may be helpful to compare them with each other.

One way to attempt to better understand hepatitis C and its differences to
other diseases caused by viral infections is to use mathematical models of the
evolution of the disease within a host. The basic model of viral infection
\cite{nowak00} is a system of three ordinary differential equations where the
unknowns are the numbers of uninfected cells $x$, infected cells $y$ and virus
particles $v$. It can be used to model any viral infection and any information
about the difference between viruses must be contained in the choice of
parameters and initial data. The basic model was originally applied to HIV
and there are two aspects of it which may be inappropriate for modelling
hepatitis C (infection with HCV) and hepatitis B (infection with HBV). The
first is that it includes a constant source term in the evolution equation
for uninfected cells. In the case of HIV the main target cells of the virus
are T cells and these can be replenished from production in the bone marrow.
In the case of HCV and HBV the target cells are the hepatocytes of the liver
and it is not clear that there is an external source for those. On the other
hand it has been suggested that blood cells coming from the bone marrow may
be transformed into hepatocytes \cite{petersen99} and to the authors'
knowledge this possibility has not been ruled out. The second aspect of the
basic
model which is problematic for its application to hepatitis is the fact
that the infection rate is described by a mass action term, i.e. one
proportional to $xv$ and this can lead to the conclusion that adults
should be more susceptible to hepatitis than children, something which is
not seen in reality \cite{gourley07}.
This can be avoided by replacing mass action by the
so-called standard incidence function, which is proportional to
$\frac{xv}{x+y}$. It has the mathematical inconvenience that this
expression is singular for $x+y=0$.

In \cite{nangue21} a model for
hepatitis C was introduced and some of the properties of its solutions
established. A central aim of what follows is to extend those results
on the dynamics of the model of \cite{nangue21}. Another aim is to
better understand those results in a wider context, both mathematical
and biological. The model of \cite{nangue21} is a system of five ODE
which augments a modification of the basic model 
by a simple description of virus production in the cell. The model for
virus production comes from \cite{guedj10}, where it was added to
a previous model of hepatitis C introduced in
\cite{neumann98}. The model of \cite{nangue21} includes a constant source of
uninfected cells and uses the standard incidence function.

The basic model has been extended in many directions, some of which are
mentioned in what follows. One key issue is that of the role of the
immune system in the infection. The immune system is not included explicitly
in the basic model. It is included implicitly due the fact that the parameters
describing the elimination of infected cells and virus particles 
may include contributions due to effects of the immune system. This issue is
important since the differences between the effects of HCV, HBV and HIV
probably have a lot to do with differences in the immune response. It is not
simple to identify these
since the mechanisms involved may go beyond the most obvious candidates.
For instance it is observed that in the initial phase of an infection with
hepatitis C the amount of virus reaches a maximum and decreases again. If
this was due to killing of infected cells an increase in liver enzymes should
be seen. In fact no such effect is observed. It is believed that the decrease
of the virus in this phase is due to the influence of natural killer cells
(NK cells). However this influence takes place not through the killing which
gives NK cells their name but due to the fact that they produce
interferon $\gamma$ \cite{missale02}. In this context it should be noted that
important features of the dynamics of
hepatitis C evolve on very different time scales, from a few days for the
acute infection to many years for the chronic infection. Thus it might be
appropriate to use different models for different phases of the disease. In
view of this the strategy of the present paper is to understand the dynamics
of the model without any restrictions on the parameters other than their
positivity. Restrictions on the parameters which are appropriate for specific
applications of the model and the restrictions on the phenomenology which may
result are left for future investigation.

In this paper we extend the analysis in \cite{nangue21} of the model introduced
there. Section \ref{intro} is concerned with basic properties of that model.
Theorem 2 states that there are up to three steady states $E_0$, $E_0'$ and
$E_0''$ on the boundary of the positive orthant. The solution $E_0''$ was not
seen in \cite{nangue21} since it is not consistent with the stronger
restrictions on the parameters made in that paper. Theorem 3 shows, among
other things, that there is an open set of initial data leading to solutions
which converge to $E_0''$ as $t\to\infty$ and that the study of the late-time
behaviour of solutions which do not converge to either $E_0$ or $E_0''$ can
be reduced to that of the late-time behaviour of a three-dimensional system.
In Section \ref{linear} the stability properties of the boundary steady states
are studied by linearization. In \cite{nangue21} a condition on the parameters
in the system was found which ensures that all solutions converge to steady
states but the question whether this is true for all values of the parameters
was left open. In Section \ref{osc} we provide a
negative answer to this question by proving that there are parameter choices
for which periodic solutions exist. This is done with the help of Hopf
bifurcations. In Section \ref{ss} a parametrization of all steady states is
given which extends one given in \cite{nangue21}.
It follows that there are at most three steady states for
any choice of parameters. In many models related to infectious diseases we
have the situation that the number of positive steady states is controlled by
a basic reproductive number $R_0$. It is zero for $R_0\le 1$ and one for
$R_0>1$. For the model of \cite{nangue21} we are able to prove that for a
certain quantity ${\cal R}_0''$ the number of steady states is even for
${\cal R}_0''\le 1$ and odd for ${\cal R}_0''>1$. We were, however, not able to
decide whether the number of positive steady states is ever greater than the
minimum consistent with this parity statement. One way in which a greater
number of positive steady states might occur is through the occurrence of
a backward bifurcation from the infection-free steady state $E_0'$, where an
unstable positive steady state arises \cite{dushoff98}, \cite{li14}.
We show that this does not happen in the model
studied here. Instead there is a forward bifurcation where a stable positive
steady state arises. We also prove the existence of a stable steady state in
a different region of parameter space by a perturbation analysis of the limit
where the source strength $s$ tends to zero. Section \ref{outlook} gives an
outlook on possible future research directions.

\section{Basic properties of the model}\label{basic}

The following model for the dynamics of the infection of a host by the
hepatitis C virus (HCV) was introduced in \cite{nangue21}.
\begin{eqnarray}
  &&\frac{dT}{dt}=s+r_TT\left(1-\frac{T+I}{T_{\rm max}}\right)-dT
     -\frac{bTV}{T+I}\label{basic1},\\
  &&\frac{dI}{dt}=r_II\left(1-\frac{T+I}{T_{\rm max}}\right)
  +\frac{bTV}{T+I}-\delta I,\label{basic2}\\
  &&\frac{dV}{dt}=\rho RI-cV-\frac{bTV}{T+I},\label{basic3}\\
  &&\frac{dU}{dt}=\beta R\left(1-\frac{U}{U_{\rm max}}\right)-\gamma U,
     \label{basic4}\\
  &&\frac{dR}{dt}=\alpha (1-\epsilon)U-\sigma R.\label{basic5}
\end{eqnarray} 
In that paper several properties of the solutions of this model were determined.
In what follows these results will be extended and some features of solutions of
this model will be compared with those of solutions of related models.

In these equations $T$, $I$, $V$, $U$ and $R$ are functions of time $t$ which
are supposed to be non-negative. All other quantities in these equations are
positive constants. $U$ and $R$ are the total amounts of certain types of
RNA in the cells and they give a representation of the replication
machinery of HCV. $T$, $I$ and $V$ are the numbers of uninfected cells,
infected cells and virus particles inside the cells. Note that the equations
(\ref{basic4})-(\ref{basic5}) form a closed system for $U$ and $R$.

The right hand sides of (\ref{basic1})-(\ref{basic5}) define smooth functions
on the positive orthant and thus local existence holds for the initial value
problem with positive initial data. These functions cannot all be extended
continuously to the non-negative orthant and so to go beyond local existence
it is important to prevent a solution which is initially positive approaching
the singular set defined by $T=I=0$. The following theorem collects some
results proved in \cite{nangue21} and extends them slightly by removing
some restrictions on the parameters.

\noindent
{\bf Theorem 1} Let $T_0,I_0,V_0,U_0,R_0$ be positive real numbers. Then there
exists a unique positive solution $(T(t),I(t),V(t),U(t),R(t))$ of
(\ref{basic1})-(\ref{basic5}) on the interval $[0,\infty)$ with $T(0)=T_0$,
$I(0)=I_0$, $V(0)=V_0$, $U(0)=U_0$ and $R(0)=R_0$. It is bounded and there
exists a constant $C>0$ such that $T+I\ge C$.

\noindent
{\bf Proof} In \cite{nangue21} it was proved that for any solution with
positive initial data defined on an interval $[0,t_1)$ with $t_1$ finite or
infinite there exists a constant $C>0$ such that $T+I\ge C$.  It was also
proved in \cite{nangue21} that all solutions of the system considered there
exist for all positive times and are bounded under the restrictions that
$d\le\delta$ and $r_I\le r_T$. In fact it is not hard to extend the proof
given there to general positive parameters. To do so let $r_1=\max\{r_T,r_I\}$
and $d_1=\min\{d,\delta\}$. Then
\begin{equation}
\frac{d(T+I)}{dt}\le s+r_1(T+I)\left(1-\frac{T+I}{K}\right)-d_1(T+I)\nonumber.
\end{equation}
This allows us to argue as in \cite{nangue21} to show that $T+I$ is bounded
and thus that the whole solution is bounded. The bound of $T+I$ by the quantity
$p_0$ introduced in \cite{nangue21} is replaced by a bound in terms of the
quantity obtained from $p_0$ by replacing $d$ by $d_1$ and $r_T$ by $r_1$.
The definition of $p_0$ is recalled in (\ref{p0}). Once the solution is known
to be bounded and remain away from the singular set it is straightforward to
show that all concentrations remain positive as long as the solution exists.
$\blacksquare$

Note that it is essential for this result that $s>0$. It was shown in
\cite{hews10} and \cite{hews21} that for some models related to the system
(\ref{basic1})-(\ref{basic5}) with $s=0$ there are solutions which tend to
the origin as $t\to\infty$. In fact it is also true, and simpler to prove,
that there are solutions of this type in the model of Guedj and Neumann
\cite{guedj10}. There the origin is a steady state and the linearization of
the system about that point has
a negative eigenvalue when $r<d$. The analogues for the system of
\cite{guedj10} of all statements in Theorem 1 other than that concerning
boundedness below by a positive constant hold and can be proved in the same way.

In order to obtain insights into the dynamics of the model a useful first
step is to determine the boundary steady states, i.e. the non-negative steady
states for which at least one of the variables vanishes.  Some information on
these is collected in the following theorem.

\noindent
{\bf Theorem 2} For given values of the parameters the system
(\ref{basic1})-(\ref{basic5}) has up to three steady states on the boundary.
Let
\noindent
\begin{equation}\label{p0}
  p_0=\left(1-\frac{d}{r_T}+\left(\left(1-\frac{d}{r_T}\right)^2
      +\frac{4s}{r_T T_{\rm max}}\right)^{\frac12}\right)
  \frac{T_{\rm max}}{2}.
\end{equation}

\noindent
1. $E_0=(p_0,0,0,0,0)$ is a non-negative steady state

\noindent
2. Let $U^*=U_{\rm max}\left(1-\frac{1}{{\cal R}_0'}\right)$ and
$R^*=U_{\rm max}\frac{\gamma}{\beta}({\cal R}_0'-1)$, where
${\cal R}_0'=\frac{\alpha\beta (1-\epsilon)}{\gamma\sigma}$.
Then if ${\cal R}_0'>1$ the point $E_0'=(p_0,0,0,U^*,R^*)$ is a
non-negative steady state.

\noindent
3. If $d>\frac{r_T\delta}{r_I}$ and
$\frac{s}{r_TT_{\rm max}}<\left(1-\frac{\delta}{r_I}\right)
\left(\frac{d}{r_T}-\frac{\delta}{r_I}\right)$ let
$T^*=s\left(d-\frac{r_T\delta}{r_I}\right)^{-1}$ and
\begin{equation}
  I^*=T_{\rm max}\left(1-\frac{\delta}{r_I}
    -\frac{s}{r_TT_{\rm max}}\left(\frac{d}{r_T}
      -\frac{\delta}{r_I}\right)^{-1}\right).\nonumber
\end{equation}
Then the point $E_0''=(T^*,I^*,0,0,0)$ is a non-negative steady state. There
are no other non-negative steady states than those given in points 1.-3.

\noindent
{\bf Proof} For a non-negative steady state of (\ref{basic1})-(\ref{basic5})
it follows from (\ref{basic1}) that $T$ is positive. If $V$ is positive then it
follows from (\ref{basic3}) that $I$ and $R$ are positive. It then follows from
(\ref{basic5}) that $U$ is positive and that we do not have a boundary steady
state. Thus boundary steady states of (\ref{basic1})-(\ref{basic5}) satisfy
$V=0$. It then follows from (\ref{basic3}) that at least one of $R$ and $I$
must be zero. The case $I=0$ was analysed completely in \cite{nangue21}.
Under the assumptions made on the parameters in \cite{nangue21} it was the
only case. In that case $T=p_0$ and only the steady states $E_0$ and $E_0'$
occur. $(U^*,R^*)$ is the unique positive steady state of the system
(\ref{basic4})-(\ref{basic5})
It remains to treat the case that $I\ne 0$, so that $R=0$. Then it
follows that $U=0$. As a consequence of (\ref{basic2}) we have
$1-\frac{T+I}{T_{\rm max}}=\frac{\delta}{r_I}$. Substituting this back into
(\ref{basic1}) gives $s=\left(d-\frac{r_T\delta}{r_I}\right)T$. If
$d\le\frac{r_T\delta}{r_I}$ this gives a contradiction. Suppose then that
$d>\frac{r_T\delta}{r_I}$. Solving for $T$ we get the equation
$T=s\left(d-\frac{r_T\delta}{r_I}\right)^{-1}$. The evolution equation for
$I$ gives $I=T_{\rm max}\left(1-\frac{\delta}{r_I}-\frac{T}{T_{\rm max}}\right)$.
Hence
\begin{equation}
  I=T_{\rm max}\left(1-\frac{\delta}{r_I}
    -\frac{s}{r_TT_{\rm max}}\left(\frac{d}{r_T}
      -\frac{\delta}{r_I}\right)^{-1}\right).\nonumber
\end{equation}
It may be concluded that necessary and sufficient conditions for the existence
of a boundary steady state with $I>0$ are $d>\frac{r_T\delta}{r_I}$ and
$\frac{s}{r_TT_{\rm max}}<\left(1-\frac{\delta}{r_I}\right)
\left(\frac{d}{r_T}-\frac{\delta}{r_I}\right)$. When these conditions are
satisfied we get explicit expressions for $T$ and $I$ and $V=R=U=0$.
$\blacksquare$

\noindent
{\bf Remark 1} In the model of \cite{guedj10} there exist boundary steady states
corresponding to $E_0$ and $E_0'$ when $d<r$, with $p_0$ replaced by
$1-\frac{d}{r}$. When $r\le d$ the only boundary steady states satisfy
$T=I=V=0$.

We recall some facts about the late time behaviour of solutions of
(\ref{basic1})-(\ref{basic5}) proved in \cite{nangue21}.

\noindent
{\bf Theorem 3} Any solution of the system (\ref{basic1})-(\ref{basic5})
with positive initial data belongs to one of the following three cases.

\noindent
1. The solution converges to $E_0$ as $t\to\infty$

\noindent
2. The solution converges to $E_0''$ as $t\to\infty$

\noindent
3. The solution passing through any $\omega$-limit point of the original 
solution is of the form $(T,I,V,U^*,R^*)$ where $(T,I,V)$ is a solution of
the system (\ref{basic1})-(\ref{basic5}) with $R=R^*$

\noindent
{\bf Proof} Given a positive solution of (\ref{basic1})-(\ref{basic5}) we can
consider a solution passing through one of its $\omega$-limit points. In the
limiting solution $R$ and $U$ are constant while the other unknowns satisfy
the equations obtained from (\ref{basic1})-(\ref{basic3}) by setting $R$ equal
to that given value, either $R=0$ or $R=R^*$. Consider first the case of the
system (\ref{basic1})-(\ref{basic3}) with $R=0$. In that case $V$ is a Lyapunov
function and $V$ converges to zero for all solutions. Considering again
a solution passing through an $\omega$-limit point leads to a solution
of the system (\ref{basic1})-(\ref{basic2}) with $V=0$. As remarked in
\cite{nangue21} this system is competitive and so each solution converges
to a steady state. In \cite{nangue21} it was claimed that all points of this
system with $T=0$ were steady states. This is not true. In fact on that
boundary we have $\dot T=c>0$ and so no positive solution can approach a
point of that boundary. On the boundary $I=0$ the unique steady state is
given by $T=p_0$. Under the assumptions made on the parameters in
\cite{nangue21} there did not exist an interior steady state. With the
less restrictive assumptions made here there is an interior steady state for
certain choices of parameters. It corresponds to the point $E_0''$
considered above. We can conclude that any solution with the property that
$R\to 0$ for $t\to\infty$ either converges to $E_0$ or $E_0''$. This completes
the proof. $\blacksquare$

Under the assumptions made on the parameters in \cite{nangue21} only the first
and third cases listed in Theorem 3 occur. It will be seen in the next section
that there are parameter values for which the second case occurs. Theorem 3
implies that either a solution has very simple asymptotics (case 1 or 2) or
the analysis of its asymptotics can be reduced to that of a solution of
(\ref{basic1})-(\ref{basic3}) with $R=R^*$. A similar statement holds for the
model defined by the equations (3) of \cite{guedj10}. Either a solution
converges to $E_0'$, it converges to a steady state with $T=I=V=0$ or its
asymptotics are determined by those of a solution of the equations (1) of
\cite{guedj10}.

\section{Linearization}\label{linear}

We now consider the linearization of the system (\ref{basic1})-(\ref{basic5})
about an arbitrary steady state. It has a block upper triangular form with the
bottom left $2\times 3$ submatrix being zero and so its eigenvalues are the
union of those of the top left $3\times 3$ submatrix and the bottom right
$2\times 2$ submatrix. The latter is the linearization of the system
(\ref{basic4})-(\ref{basic5}) and is of the form
\begin{equation}
\left[
{\begin{array}{cc}\label{lin2}
-\frac{\beta R}{U_{\rm max}}-\gamma & \beta\left(1-\frac{U}{U_{\rm max}}\right) \\
\alpha (1-\epsilon) & -\sigma   
\end{array}}
\right].
\end{equation}
The former is the linearization of the system (\ref{basic1})-(\ref{basic3})
and is of the form
\begin{equation}
\left[
{\begin{array}{ccc}\label{lin3}
   -a_1 & b_1 & -\frac{bT}{T+I}\\
   a_2 & -b_2 & \frac{bT}{T+I}\\
   -\frac{bIV}{(T+I)^2} & \rho R+\frac{bTV}{(T+I)^2} & -c-\frac{bT}{T+I}
\end{array}}
\right]\nonumber
\end{equation}     
where
\begin{eqnarray}
  &&a_1=d-r_T+\frac{2r_TT}{T_{\rm max}}+\frac{r_TI}{T_{\rm max}}
  +\frac{bIV}{(T+I)^2},\nonumber\\
  &&a_2=-\frac{r_II}{T_{\rm max}}+\frac{bIV}{(T+I)^2},\nonumber\\
  &&b_1=-\frac{r_TT}{T_{\rm max}}+\frac{bTV}{(T+I)^2},\nonumber\\
  &&b_2=\delta-r_I+\frac{r_IT}{T_{\rm max}}+2\frac{r_II}{T_{\rm max}}
+\frac{bTV}{(T+I)^2}.\nonumber    
\end{eqnarray}

\noindent
{\bf Theorem 4} The eigenvalues of the linearization at the boundary steady
states of the system (\ref{basic1})-(\ref{basic5}) have the following
properties.

\noindent
1. The number of eigenvalues at the point $E_0$ which have positive real parts
is equal to the number of positive quantities among  
$\left(1-\frac{\delta}{r_I}\right)\left(\frac{d}{r_T}-\frac{\delta}{r_I}\right)
-\frac{s}{r_IT_{\rm max}}$ and ${\cal R}_0'-1$. 

\noindent
2. The point $E_0'$ always has at least four eigenvalues with negative real
part. The sign of the remaining eigenvalue is equal to that of ${\cal R}_0''-1$.

\noindent
3. All eigenvalues of $E_0''$ have negative real part.

\noindent
{\bf Proof} It was shown in \cite{nangue21} that the eigenvalues of
(\ref{lin2}) at the origin have negative real parts for ${\cal R}_0'<1$ and
that for ${\cal R}_0'>1$ there is one positive and one negative eigenvalue.
It was also shown that the eigenvalues of (\ref{lin2}) at the positive steady
state always have negative real parts. 

If we evaluate the matrix (\ref{lin3}) at the point $E_0$ we see that the
entries below the diagonal vanish so that the eigenvalues are just the
diagonal elements. The third diagonal element is obviously negative. The
first is $r_T\left[\left(1-\frac{d}{r_T}\right)-\frac{2p_0}{T_{\rm max}}\right]$,
which is also negative. The sign of the second depends on the choice of
parameters. Note that $p_0$ is a positive root of a quadratic polynomial $p$
which has a positive leading term and is negative at the origin. The eigenvalue
has the same sign as the value of the polynomial at the point
$T_{\rm max}\left(1-\frac{\delta}{r_I}\right)$. Hence it is negative precisely 
when
\begin{equation}\label{Yineq}
  \left(1-\frac{\delta}{r_I}\right)\left(\frac{d}{r_T}-\frac{\delta}{r_I}\right)
  <\frac{s}{r_IT_{\rm max}}.
\end{equation}
This proves 1.

The signs of the real parts of the eigenvalues of the linearization at the
point $(p_0,0,0,R^*,U^*)$ were determined in \cite{nangue21} and this proves 2.

If we evaluate the matrix (\ref{lin3}) at the point $E_0''$ we get
\begin{equation}
\left[
{\begin{array}{ccc}
   r_T-d-\frac{r_T(2T+I)}{T_{\rm max}} &
   -\frac{r_TT}{T_{\rm max}} & -\frac{bT}{T+I}\\
   -\frac{r_II}{T_{\rm max}}& r_I-\delta-\frac{r_I(T+2I)}{T_{\rm max}} 
 & \frac{bT}{T+I}\\
   0 & 0 & -c-\frac{bT}{T+I}
\end{array}}
\right].\nonumber
\end{equation}
After some manipulation using the expressions for $T$ and $I$ at the steady
state it can be shown that the trace of this matrix is equal to
$-d+\frac{r_T\delta}{r_I}-\frac{r_TT}{T_{\max}}-\frac{r_II}{T_{\max}}<0$.
Its determinant is given by
$\frac{r_II}{T_{\max}}\left(d-\frac{r_T\delta}{r_I}\right)
+\frac{r_Tr_ITI}{T_{\max}^2}>0$. We conclude that all eigenvalues have negative
real part. This proves 3. $\blacksquare$ 

\noindent
{\bf Remark 2} It follows from 3. that $E_0''$ is a hyperbolic sink. In
particular this implies that there exist positive solutions which converge to
$E_0''$ as $t\to\infty$.

\noindent
{\bf Remark 3} If the parameters are varied so that the quantities on the left
and right hand sides of (\ref{Yineq}) become equal then $I$ tends to zero,
which means that $E_0$ and $E_0''$ approach each other.

\noindent
{\bf Remark 4} In the system of \cite{guedj10} the number of positive
eigenvalues at the point $E_0$ is equal to the number of positive quantities
among $\frac{d}{r}-1$ and ${\cal R}_0'$. The stability properties of $E_0'$
are as in Theorem 4. The number of positive eigenvalues at one of the remaining
steady states is zero for ${\cal R}_0'<1$ and one for ${\cal R}_0'>1$

Consider now the case where $E_0''$ does not exist. Then all solutions must
approach $E_0$ and this suggests that $-b_2<0$. One case in which $E_0''$ does
not exist is when $d\le\frac{r_T\delta}{r_I}$. Then
$\frac{d}{r_T}\le\frac{\delta}{r_I}$. We will show that this indeed implies
that $-b_2<0$. If $\frac{d}{r_T}\ge 1$ then $\frac{\delta}{r_I}\ge 1$ and
$-b_2<0$. If, on the other hand, $\frac{d}{r_T}<1$ then
$p_0>\left(1-\frac{d}{r_T}\right)T_{\rm max}
\ge \left(1-\frac{\delta}{r_I}\right)T_{\rm max}$ and again $-b_2<0$. The other
case where $E_0''$ does not exist is that where $d>\frac{r_T\delta}{r_I}$ and
$\frac{s}{r_TT_{\rm max}}\ge\left(1-\frac{\delta}{r_I}\right)
\left(\frac{d}{r_T}-\frac{\delta}{r_I}\right)$. It has been shown above that
when this inequality is strict $-b_2<0$ while in the case of equality $b_2=0$.

%We may hypothesize that
%$-b_2<0$ in this case too. First consider the situation where 
%$\frac{s}{r_TT_{\rm max}}=\left(1-\frac{\delta}{r_I}\right)
%\left(\frac{d}{r_T}-\frac{\delta}{r_I}\right)$. Then at the point $E_0''$
%we have $\frac{T}{r_TT_{\rm max}}=\frac{s}{r_TT_{\rm max}}
%\left(d-\frac{r_T\delta}{r_I}\right)^{-1}$.

\section{Oscillations}\label{osc}

It has been known for a long time that in-host models of viral dynamics can
exhibit oscillations. This was seen numerically in a model for HIV in
\cite{perelson93}. The model studied in that paper has four unknowns,
since it distinguishes between latently and productively infected cells.
It uses mass-action kinetics for the process of infection. There is both
a constant source and a logistic proliferation of uninfected cells. It was
found that for certain values of the parameters there are periodic solutions
which arise in a Hopf bifurcation. However these parameter values do not seem
to be appropriate for the concrete application considered in \cite{perelson93}.
A version of this model without latently infected cells was studied
mathematically in \cite{deleenheer03}. It was proved that there exist stable
periodic solutions for certain values of the parameters. This model is similar
to that of \cite{guedj10} but not identical to it. The difference is that
the expression $T+I$ in the model of \cite{guedj10} is replaced by $T$.
This is essential for the proofs in \cite{deleenheer03} since they make use
of the fact that the system is three-dimensional and competitive.

In this section it will be proved that the system (\ref{basic1})-(\ref{basic5})
possesses periodic solutions and thus, in particular, that there are solutions
which do not converge to a steady state as $t\to\infty$. In order to do this
we start with the system obtained from (\ref{basic1})-(\ref{basic3}) by setting
$R=R^*$, $s=0$, $d=0$ and $r_I=0$ and replacing the last term in
(\ref{basic3}) by $-\frac{\eta bTV}{T+I}$ with a constant $\eta$. The result is
\begin{eqnarray}
  &&\frac{dT}{dt}=r_TT\left(1-\frac{T+I}{T_{\rm max}}\right)
     -\frac{bTV}{T+I}\label{basic31},\\
  &&\frac{dI}{dt}=\frac{bTV}{T+I}-\delta I,\label{basic32}\\
  &&\frac{dV}{dt}=\rho R^*I-cV-\frac{\eta bTV}{T+I}.\label{basic33}
\end{eqnarray}
It is identical to a model studied in \cite{hews10} except for an additional
term in the last equation which takes account of the absorption of virions by
the cells they infect and the use of a different notation. It coincides with 
the system of \cite{hews10} in the case $\eta=0$ while the case $\eta=1$ is
that with absorption of virions. The equations are defined and smooth except
when $T+I=0$. There are at most two steady states in the regular region
where $T+I\ne 0$. The first is a disease-free steady state with coordinates
$(p_0,0,0)$. The second, when it exists, has coordinates
\begin{eqnarray}
  &&T^*=\frac{T_{\max}\delta}{r_T}\left(\frac{R_1}{R_0}-1\right)\label{ss1},\\
  &&I^*=\frac{T_{\max}\delta}{r_T}\left(\frac{R_1}{R_0}-1\right)(R_0-1),\label{ss2}\\
  &&V^*=\frac{T_{\max}\delta\rho R^*\omega}{r_T}\left(\frac{R_1}{R_0}-1\right)(R_0-1),
     \label{ss3}
\end{eqnarray}
where $R_0=\frac{b}{c}\left(\frac{\rho R^*}{\delta}-\eta\right)$, 
$R_1=\frac{r_Tx+\delta}{\delta}$ and
$\omega=\left(c+\frac{\eta b}{R_0}\right)^{-1}$.
These formulae generalize those given for the case $\eta=0$ in \cite{hews10}.
We see that a positive steady state exists precisely when $1<R_0<R_1$. Let
$X=\frac{T^*}{T^*+I^*}$. Then
$X=R_0^{-1}=\frac{c\delta}{b(\rho R^*-\eta\delta)}$. If we let
$X_-=(R_1)^{-1}=\frac{\delta}{r_T+\delta}$ then the values of $X$ corresponding
to steady states lie in the interval $(X_-,1)$.

The linearization at the positive steady state is of the form
\begin{equation}
\left[
  {\begin{array}{ccc}
     r_T\left(1-\frac{(R_0+1)T^*}{T_{\rm max}}\right)
     -\frac{b\rho R^*\omega (R_0-1)^2}{R_0^2}&
     -\frac{r_TT^*}{T_{\rm max}}+\frac{b\rho R^*\omega (R_0-1)}{R_0^2}
     & -\frac{b}{R_0}\\
     \frac{b\rho R^*\omega (R_0-1)^2}{R_0^2}     &
     -\frac{b\rho R^*\omega (R_0-1)}{R_0^2}-\delta
     &  \frac{b}{R_0}\\
     -\frac{\eta b\rho R^*\omega (R_0-1)^2}{R_0^2}&
       \rho R^*+\frac{\eta b\rho R^*\omega (R_0-1)}{R_0^2}&
     -c-\frac{\eta b}{R_0}\\
   \end{array}}
\right].\nonumber
\end{equation}

The eigenvalues are the roots of the characteristic polynomial
$\lambda^3+a_2\lambda^2+a_1\lambda+a_0$ where
\begin{eqnarray}
  &&a_2=c+\frac{\delta R_1+\eta b}{R_0},\nonumber\\
  &&a_1=2\delta^2\left(\frac{R_1}{R_0}-1\right)(R_0-1)
     +\frac{\delta^2}{R_0^2}(R_0-1)^2
     -\frac{\delta r_T}{R_0}(R_0-1)+\delta c\left(\frac{R_1}{R_0}-1\right)
     \nonumber\\
  &&+\frac{\eta b^2\rho R^*\omega (R_0-1)}{R_0^2}
     +\frac{\eta \delta b R_1}{R_0^2},\nonumber\\
  &&a_0=\delta^2c (R_0-1)\left(\frac{R_1}{R_0}-1\right)
     -\eta \delta^2 b (R_0-1)\nonumber\\
  && +\frac{\eta b}{R_0}\left[\delta^2\left(\frac{R_1}{R_0}-1\right)
     +\delta(R_0-1)
     +\frac{\delta b\rho R^* (R_0+1)}{R_0}\left(\frac{R_1}{R_0}-1\right)\right].
     \nonumber
\end{eqnarray}

\noindent
{\bf Remark 5} Later we will be interested in the limit of these equations
where $b\to 0$, $b/c$ is constant and the parameters $r_T$, $T_{\rm max}$,
$\delta$ and $\rho R^*$ are constant. In this limit $R_0$ and $R_1$ are
constant while $\omega=O(b^{-1})$. If $a_{i,0}$ denotes the values of these
coefficients when $\eta=0$ then $a_i=a_{i,0}+O(b)$ for $b\to 0$.

\noindent
{\bf Lemma 1} There are parameter values for which the linearization of the
system (\ref{basic31})-(\ref{basic33}) with $\eta=0$ or $\eta=1$ about the
unique steady state has a pair of purely imaginary eigenvalues and the
remaining eigenvalue is negative.

\noindent
{\bf Proof} The proof uses the Routh-Hurwitz criterion. Both $a_2$ and
$a_{0,0}$ are positive. It can be ensured that $a_0$ is positive by choosing
$b$ sufficiently small as described in Remark 5. Note that if we start with 
parameters for which the condition $R_0>1$ holds, this condition is 
maintained in this limiting procedure. In
\cite{hews10} two parameters are introduced in order to understand
the eigenvalues of the linearization at the point $(T^*,I^*,V^*)$.
Replacing $\delta$ by $\hat\delta$ to avoid a conflict of notation they are
\begin{eqnarray}
  &&\hat\delta=\left[\delta^2\left(\frac{R_1}{R_0}-1\right)
     +\frac{\delta^2}{R_0^2}-\frac{\delta r_T}{R_0}\right](R_0-1)
  +\delta c\left(\frac{R_1}{R_0}-1\right),\nonumber\\
&&\sigma=\frac{-\delta^2(R_1-R_0)(R_0-1)}{R_0}.\nonumber
\end{eqnarray}
To apply the Routh-Hurwitz
criterion we must consider the quantity
\begin{equation}
a_2a_1-a_0=\left(c+\delta\frac{R_1}{R_0}\right)\hat\delta
+\delta^2\left(c+\delta\frac{R_1}{R_0}\right)
\left(\frac{R_1}{R_0}-1\right)(R_0-1)
+\eta\xi\nonumber
\end{equation}
where $\xi=O(b)$. 
As observed in \cite{hews10} in the case $\eta=0$ the linearization has a pair
of purely imaginary eigenvalues precisely when $\hat\delta=\sigma$. In the
general case the condition for imaginary eigenvalues is modified to
\begin{equation}
\hat\delta+\eta\xi\frac{R_0^2}{cR_0+\delta R_1}=\sigma.\label{imaginary}
\end{equation}
Now we prove that in the case $\eta=0$ there are parameters for which the
condition $\hat\delta=\sigma$ is satisfied. We note that no proof of this fact
was given in \cite{hews10} or in the later more general paper \cite{hews21}.
Consider fixed positive values of the parameters $\delta$, $r_T$ and $c$.
Then $R_1$ is also fixed. We can vary $R_0$ freely in the interval $(1,R_1)$,
for instance by varying $\rho$ between $\frac{\delta c}{bR^*}$ and
$\frac{(\delta+r_T)c}{bR^*}$ while keeping the other parameters fixed.
We think of $\hat\delta$ and $\sigma$ as functions of $R_0$. The quantity
$\sigma$ is zero for $R_0$ at both ends of the interval and negative in between.
Now $\hat\delta (1)=\delta^2+c r_T>0$ and
$\hat\delta(R_1)=\frac{\delta^2r_T^2}{(\delta+r_T)^2}
\left[\left(\frac{\delta}{r_T}\right)^2
  -\left(\frac{\delta}{r_T}\right)-1\right]$.
For suitable values of $r_T$ and $\delta$, for instance with $r_T=\delta$,
$\hat\delta(R_1)<0$ and so by the intermediate value theorem there is a value of
$R_0$ for which $\hat\delta=\sigma$. Consider now the case $\eta=1$.
$R_0$ can be varied between $1$ and $R_1$ by varying $\rho$ between
$\frac{\delta c}{bR^*}+1$ and $\frac{(\delta+r_T)c}{bR^*}+1$. If $b$ is
sufficiently small the quantity on the left hand side of (\ref{imaginary}) is
positive for $R_0=1$ and negative for $R_0=R_1$. The expressions for
$\hat\delta (1)$ and $\hat\delta(R_1)$ are as in the case $\eta=0$ and so in the
case $\eta=1$ (\ref{imaginary}) is satisfied at some point. $\blacksquare$ 

\noindent
{\bf Remark 6} In \cite{hews10} it is shown numerically that in the case
$\eta=0$ there is a Hopf bifurcation at points in parameter space where the
condition $\hat\delta=\sigma$ is satisfied and that periodic solutions arise.
No analytical proof of the existence of a Hopf bifurcation is given in
\cite{hews10}.

\noindent
{\bf Theorem 5} There are parameter values for which the system
(\ref{basic31})-(\ref{basic33}) with $\eta=0$ or $\eta=1$ has Hopf
bifurcations and therefore periodic solutions.

\noindent
{\bf Proof} In order to prove that a Hopf bifurcation occurs it remains to
show that
the eigenvalues which lie on the imaginary axis at the bifurcation point
cross the axis with non-zero velocity as a suitable parameter is varied
\cite{kuznetsov10}. Assuming the existence of the bifurcation point this
crossing condition has been verified for the model with $\eta=0$ in
\cite{hews21}. We will verify it in a slightly different way which will then
be modified to treat the case $\eta=1$. It follows from a result of
\cite{liu94} that in order to verify the crossing condition it is enough to
show that the Hurwitz coefficient $a_2a_1-a_0$ passes through zero with
non-zero velocity. To achieve this in the case $\eta=0$ we need a curve
$\zeta (u)$ in the space of parameters $(r_T,T_{\rm max},b,\delta,\rho R^*,c)$,
passing through the bifurcation point for $u=0$, which satisfies
$\nabla (\hat\delta-\sigma)\cdot\zeta'(0)\ne 0$. Curves of this kind exist when
the gradient of $\hat\delta-\sigma$ at the bifurcation point is non-zero. To
show that this holds we can again fix the parameters $\delta$, $r_T$ and $c$ and
vary $R_0$ freely. We have
\begin{eqnarray}
  &&\frac{d\hat\delta}{dR_0}=\frac{1}{R_0^2}[-\delta^2-c (\delta+r_T)]
     -\frac{2\delta^2}{R_0^3}\nonumber,\\
&&\frac{d\sigma}{dR_0}=\delta^2\left(1-\frac{1}{R_0^2}\right).\nonumber
\end{eqnarray}
Hence
\begin{equation}
  \frac{d(\hat\delta-\sigma)}{dR_0}=-\delta^2-\frac{c (\delta+r_T)}{R_0^2}
  -\frac{2\delta^2}{R_0^3}<0.\nonumber
\end{equation}
It follows that the condition on the gradient is always satisfied. In the
case $\eta=1$ we can argue similarly. For $b$ small the gradient of
$\hat\delta-\sigma+\eta\xi\frac{R_0^2}{c R_0+\delta R_1}$ may not be non-zero
everywhere but it will be non-zero somewhere. Since it has now been
proved that the system admits Hopf bifurcations, both for $\eta=0$ and
$\eta=1$, it follows that in both cases there are parameters for which it
admits non-constant periodic solutions \cite{kuznetsov10}. $\blacksquare$

This result can be used to prove that the system of five equations describing
the in-host dynamics of hepatitis C studied in \cite{nangue21} admits Hopf
bifurcations and hence periodic solutions for certain values of its parameters.
Consider first the system (1.3a)-(1.3c) of \cite{nangue21}. Setting the
parameters $s$, $d$ and $r_I$ in that system to zero gives, up to differences in
notation, exactly the system (\ref{basic31})-(\ref{basic33}) with $\eta=1$.
To avoid any confusion we note that, even after allowing for the differences
in notation for the basic quantities, the basic reproductive ratio $R_0$ given
above for the case $\eta=1$ is different from the quantity ${\cal R}''_0$ used
in \cite{nangue21}. In fact, as explained in \cite{vandendriessche02}, there
are choices which can be made in defining $R_0$ so that this quantity is not
unique. What is independent of these choices are the relations $R_0<1$,
$R_0=1$ and $R_0>1$. That this is true for the choices of $R_0$ made in
\cite{hews10} and \cite{nangue21} can easily be checked directly - in the
notation used above ${\cal R}''_0=\frac{b\rho R^*}{(b+c)\delta}$.

It will now be shown that the system (1.3a)-(1.3c) of \cite{nangue21} inherits
the Hopf bifurcations and hence the periodic solutions from the system
(\ref{basic31})-(\ref{basic33}) with $\eta=1$. To do this we use the following
result.

\noindent
{\bf Lemma 2} Let $f:U\times (-\epsilon,\epsilon)^{k+1}\to\R^n$,
$(x,\alpha_0,\ldots,\alpha_k)\mapsto f(x,\alpha_0,\ldots,\alpha_k)$ be a $C^1$
mapping, where $U$ is an open subset of $\R^n$. Suppose that
$f(0,0,\ldots,0)=0$, that the parameter-dependent system
$\dot x=f(x,\alpha_0,0,\ldots,0)$ has
a Hopf bifurcation at $(0,0,\ldots,0)$ and that no more than two eigenvalues of
$D_xf(0,0,\ldots,0)$ lie on the imaginary axis. Then for sufficiently small
fixed values of $(\alpha_1,\ldots,\alpha_k)$ the system
$\dot x=f(x,\alpha_0,\alpha_1,\ldots,\alpha_k)$ has a Hopf bifurcation at
$(x^*,0)$ for some $x^*(\alpha_1,\ldots,\alpha_k)$ close to zero.

\noindent
{\bf Proof} Denote the derivative $D_xf$ by $A$. Then by assumption
$A(0,0,\ldots,0)$ has a pair of purely imaginary eigenvalues and all other
eigenvalues lie off the imaginary axis. Let $\lambda^*$ be one of the
imaginary eigenvalues. Because of the smooth dependence of simple
eigenvalues on the matrix there is a $C^1$ function
$\lambda(\alpha_0,\alpha_1,\ldots,\alpha_k)$ defined on a neighbourhood of the
origin with $\lambda (0,\ldots,0)=\lambda^*$. By the definition of a Hopf
bifurcation $\frac{\partial}{\partial\alpha_0}({\rm Im}\ \lambda)\ne 0$ at the
bifurcation point. Hence for fixed $(\alpha_1,\ldots,\alpha_k)$ in a small
neighbourhood of the origin in $\R^k$ the
quantity ${{\rm Im}\ \lambda}$ has one sign for $\alpha_0$ small and positive
and the other sign for $\alpha_0$ small and negative. By the implicit function
theorem the curve $\lambda(\alpha_0,\alpha_1,\ldots\alpha_k)$ for
$(\alpha_1,\ldots,\alpha_k)$ fixed and sufficiently small also passes through
the imaginary axis for some $\alpha_0$ where its derivative is non-zero. Thus
a Hopf bifurcation occurs at that point. $\blacksquare$

\noindent
{\bf Theorem 6} There are parameter values for which the system
(\ref{basic1})-(\ref{basic5}) has Hopf bifurcations and therefore periodic
solutions.

\noindent
{\bf Proof} The system obtained from (1.3a)-(1.3c) of \cite{nangue21} by
setting $s$, $d$
and $r_I$ to zero admits a Hopf bifurcation depending on the parameter $\rho$.
Now Lemma 2 will be used to show that there are also Hopf bifurcations in the
system with $s$, $d$ and $r_I$ small and positive. It suffices to choose
$\alpha_0=\rho$, $\alpha_1=s$, $\alpha_2=d$ and $\alpha_3=r_I$. This proves
that (1.3a)-(1.3c) of \cite{nangue21} admits a Hopf bifurcation.
For a suitable choice of
parameters augmenting a periodic solution of (1.3a)-(1.3c) of \cite{nangue21}
by the constants $(U^*,R^*)$ gives a periodic solution of the full system
(1.3a)-(1.3e) of that paper. $\blacksquare$

In \cite{hews21} the authors considered a system which corresponds to
that obtained from (1.3a)-(1.3c) of \cite{nangue21} by setting $d=s=\eta=0$,
system (14)-(16) of \cite{hews21}. Lemma 2 can be used as above to prove
that that system has Hopf bifurcations for small values of the parameter
$\rho$ in that system, which corresponds to $r_I$ in our notation. In
\cite{hews21} it was suggested that there are Hopf bifurcations in that case
on the basis of computer calculations.

The proofs above give us no information about the stability of the periodic
solutions arising. The following simulation indicates that there exist
sustained oscillations. It corresponds to the parameter values $s= 10$; $d= 
10^{-5}$; $ r_{T}=0.99$; $ r_{I}= 10^{-3}$;
$T_\textrm{max}= 10 \times 10^{8}$; $b = 0.0014$; $\delta =
0.0693$; $c = 0.693$ ; $\rho = 21.5$; $ \alpha = 50$;
$\beta = 15$; $ \epsilon = 0.5 $; $\sigma = 30$; $U_\textrm{max} =
30$; $\gamma=5$ and initial data  $(T_{0}, I_{0}, V_{0}, U_{0}, V_{0})=(1000, 500, 4, 15, 20)$.
\begin{figure}[!h]
	\centering
	\begin{subfigure}[b]{0.45\textwidth}
		\includegraphics[angle=0,height=5cm,width=\textwidth]{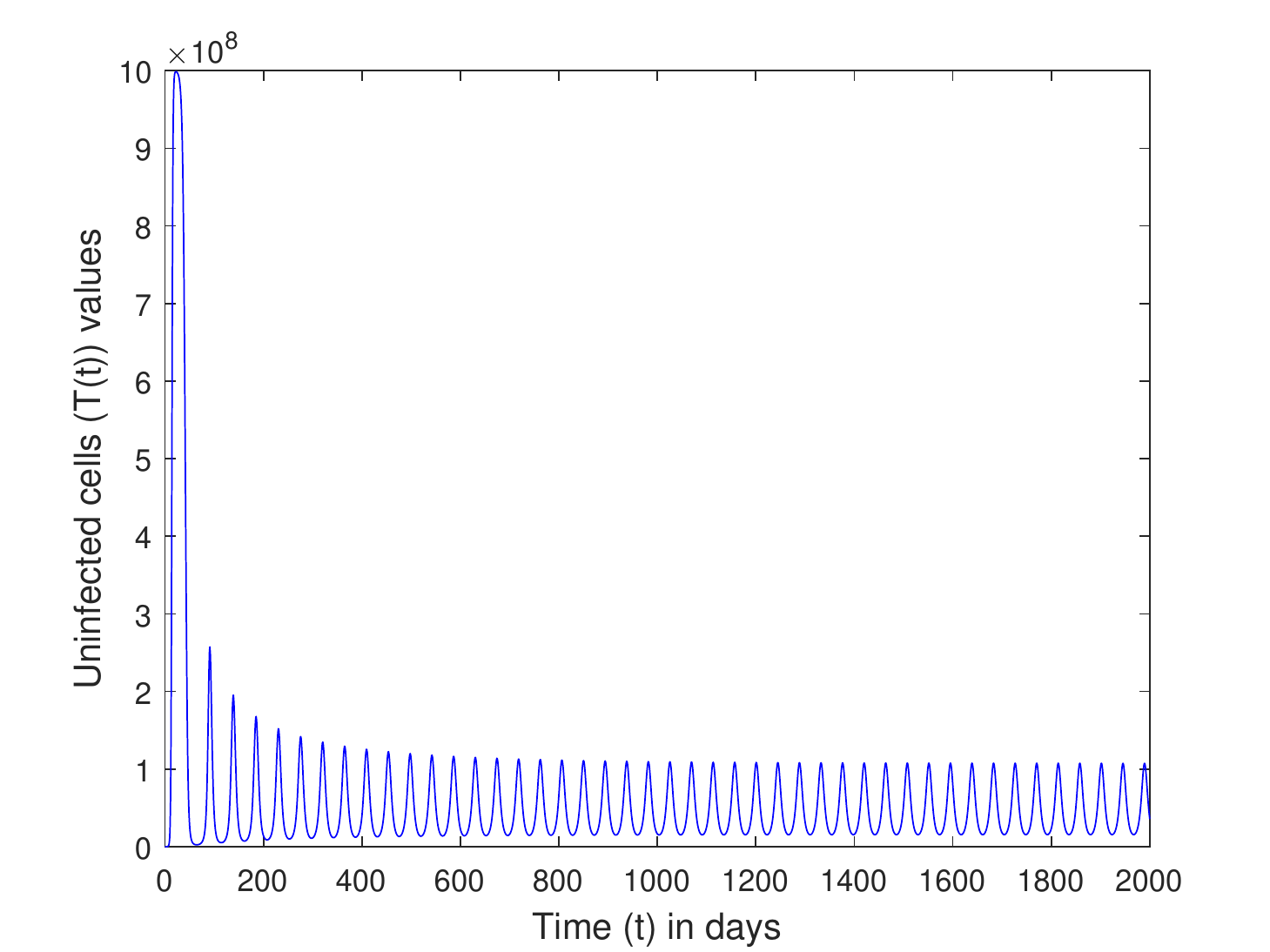}
	\end{subfigure}
	\begin{subfigure}[b]{0.45\textwidth}
		\includegraphics[angle=0,height=5cm,width=\textwidth]{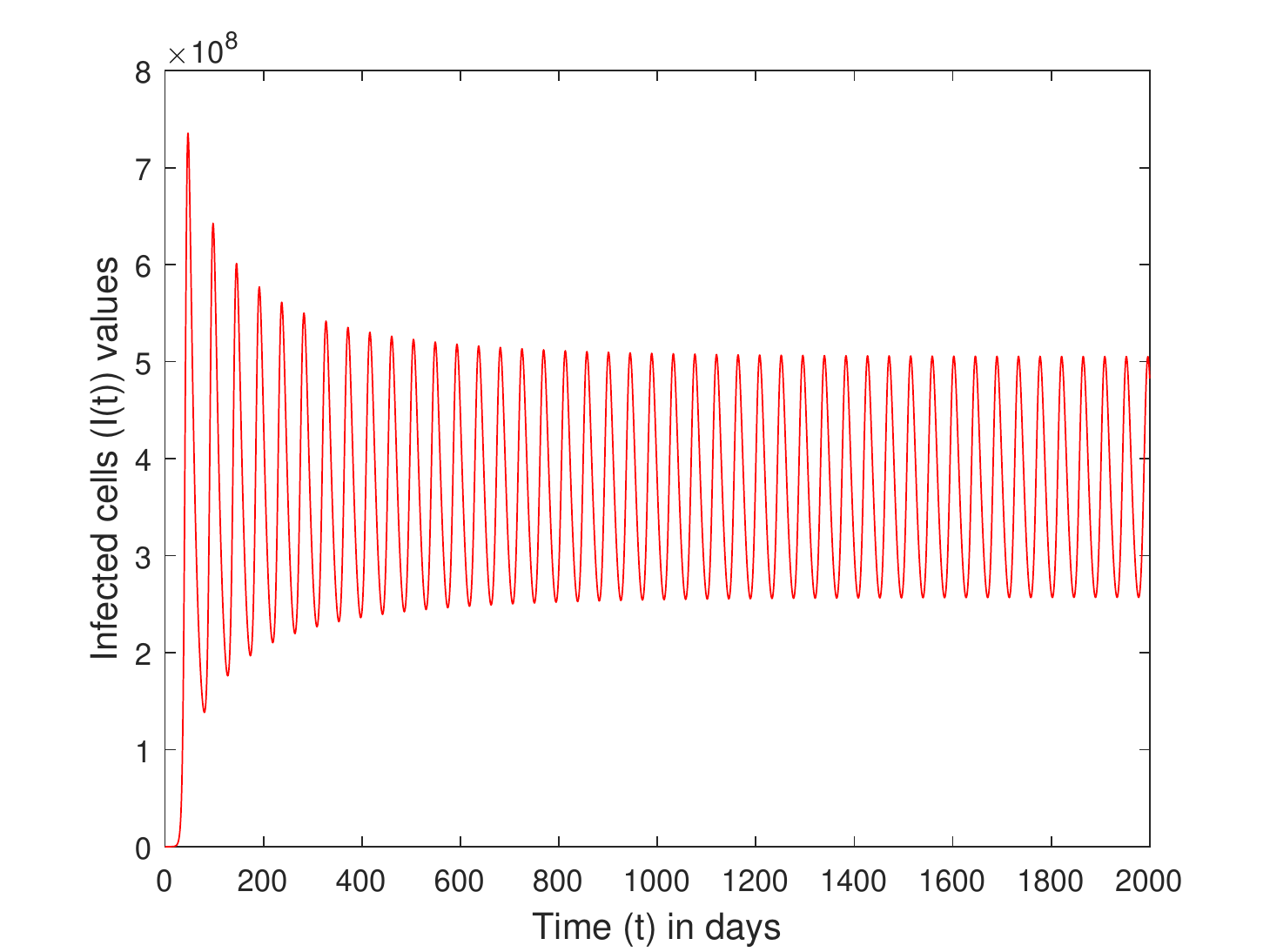}
	\end{subfigure}
	\begin{subfigure}[b]{0.45\textwidth}
		\includegraphics[angle=0,height=5cm,width=\textwidth]{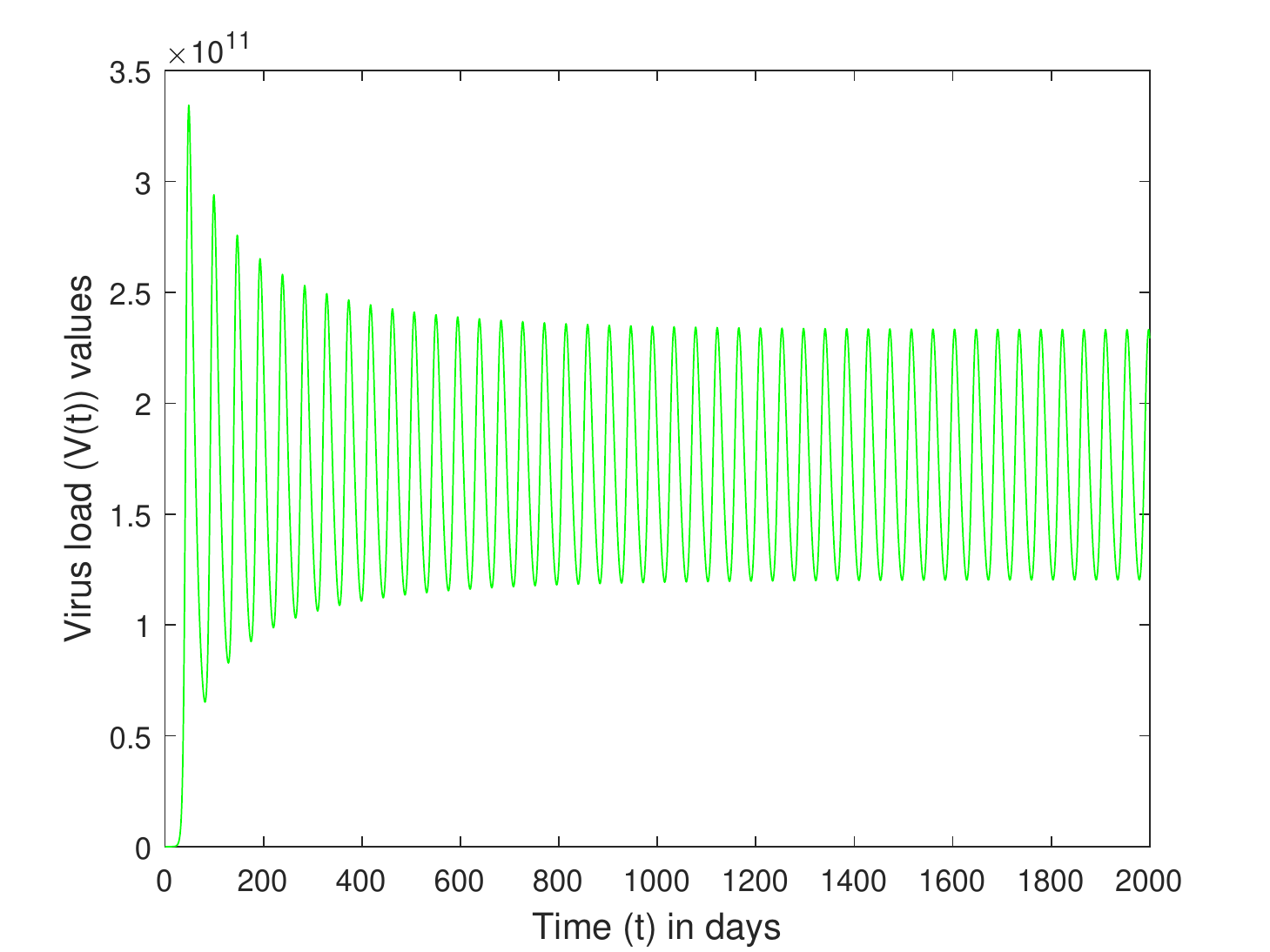}
	\end{subfigure}
	\begin{subfigure}[b]{0.45\textwidth}
		\includegraphics[angle=0,height=5cm,width=\textwidth]{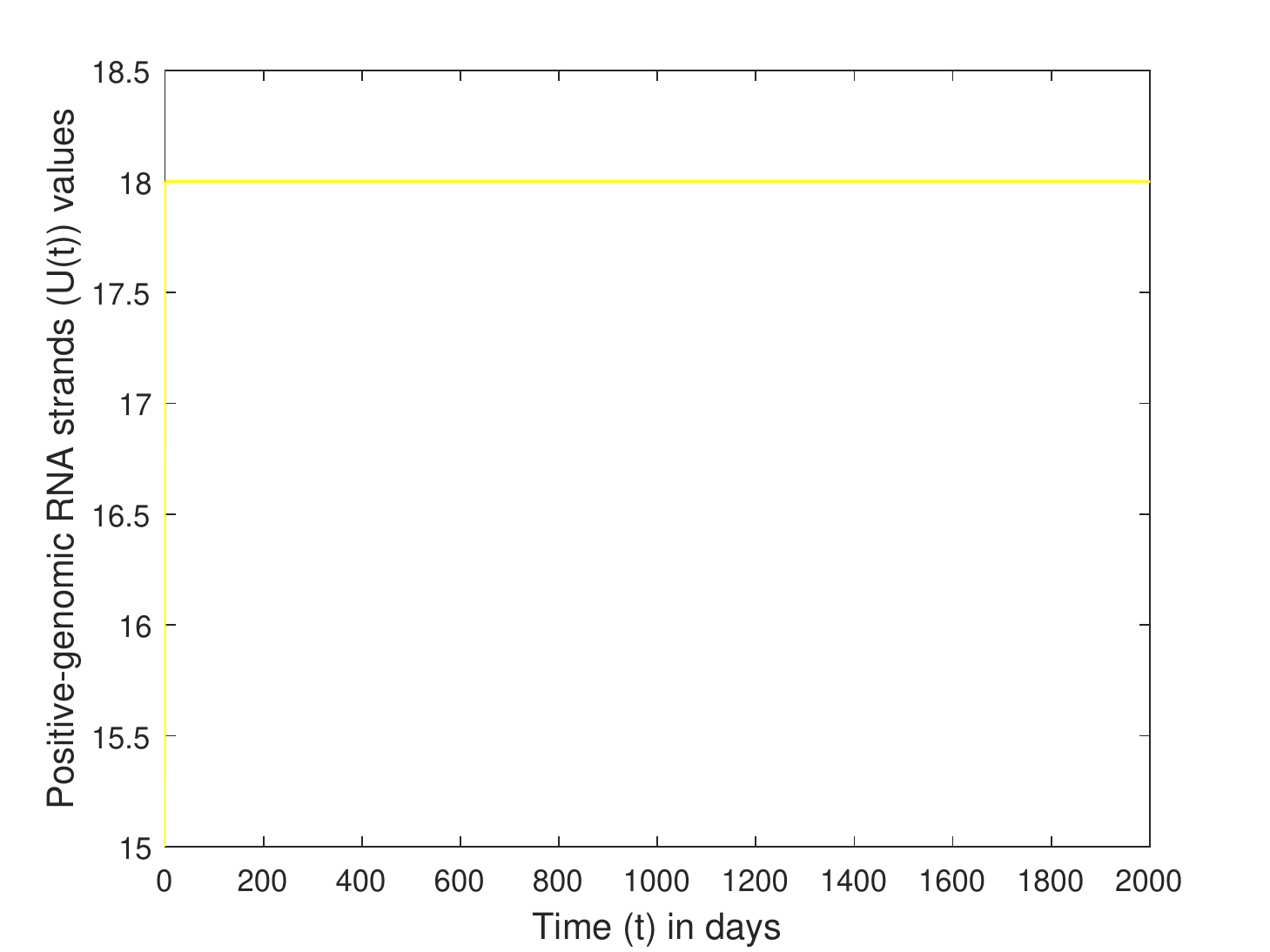}
	\end{subfigure}
	\begin{subfigure}[b]{0.40\textwidth}
		\includegraphics[angle=0,height=5cm,width=\textwidth]{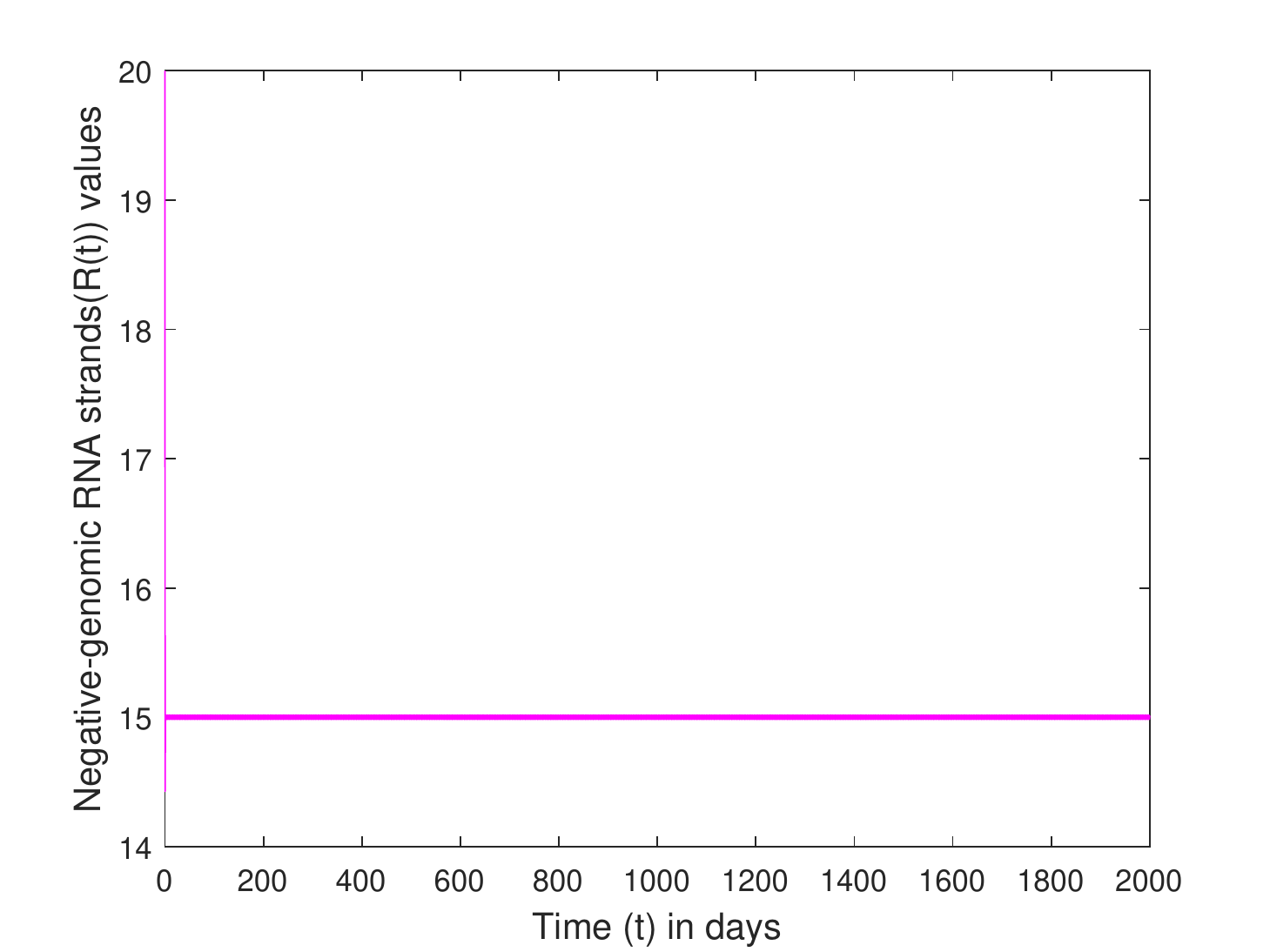}
	\end{subfigure}
	\begin{subfigure}[b]{0.5\textwidth}
		\includegraphics[angle=0,height=5cm,width=\textwidth]{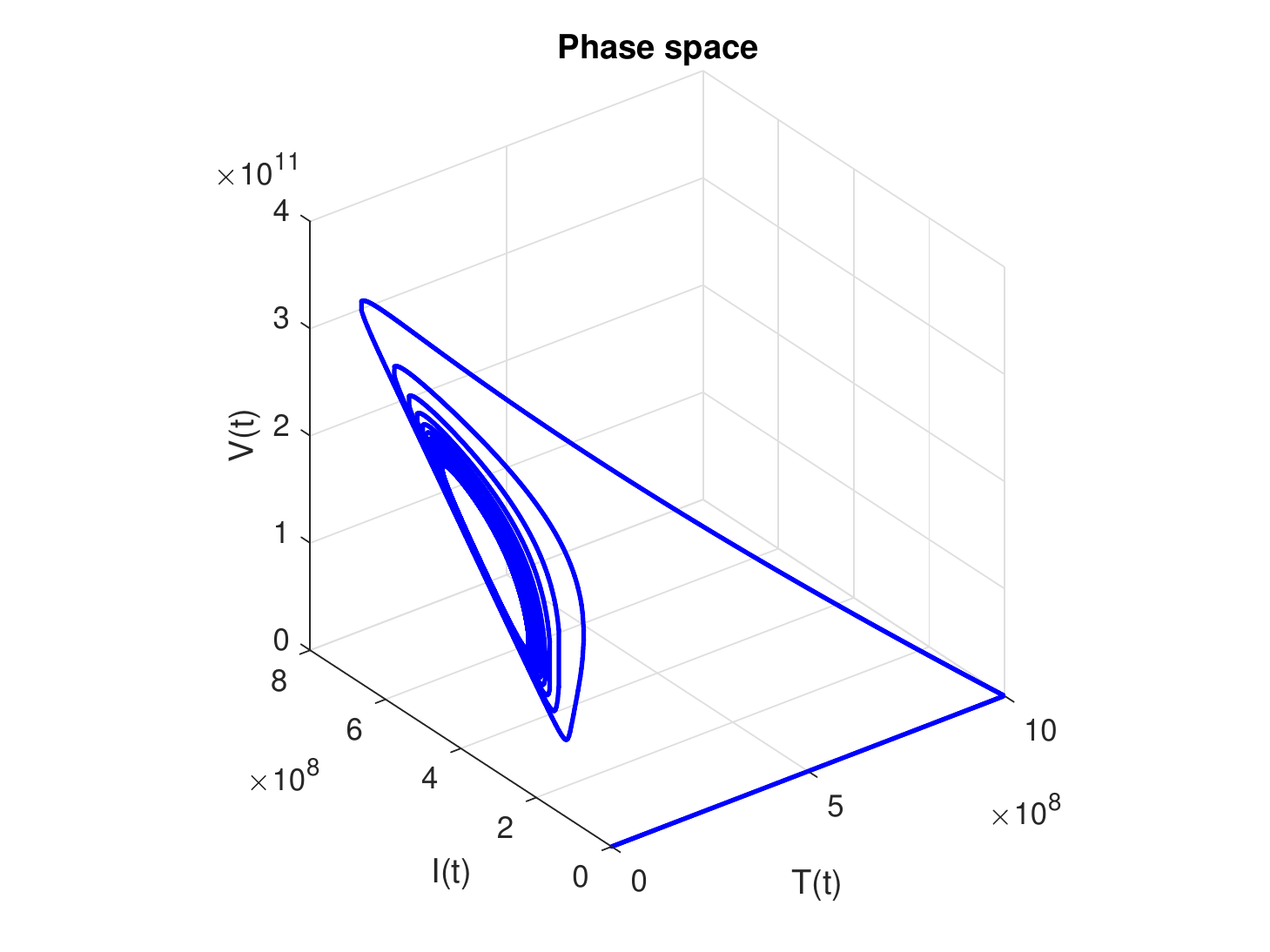}
	\end{subfigure}
	\caption{The time histories and the phase trajectory of system (\ref{basic1})-(\ref{basic5}) after Hopf bifurcation occurs for
		the above parameter values
		Such that  the composite basic
		reproductive number  $\mathcal{R}''_{0}=25.0270$ and intracellular basic reproductive number  $\mathcal{R}'_{0}=
		2.5$.} \label{fig1}
\end{figure}

\section{Positive steady states and their stability}\label{ss}

In \cite{nangue21} it was left open how many positive steady states the system
(1.3a)-(1.3c) of that paper has. It was proved that there are at most three.
However it was not decided whether there can be more than one or whether there
must be at least one when the quantity ${\cal R}''_0$ introduced in
\cite{nangue21} is greater than one. An expression for this quantity in the
notation used above is 
\begin{equation}\label{R0}
  {\cal R}''_0=\frac{b\rho R^*}{(b+c)(\delta-r_I(1-\frac{p_0}{T_{\rm max}}))},
  \nonumber
\end{equation}
where $p_0$ is the quantity defined in (\ref{p0}). The expression for
${\cal R}''_0$ in (\ref{R0}) can be obtained from
the calculation of the basic reproductive number presented in
\cite{vandendriessche02} when its denominator is positive. Otherwise it cannot
since condition (A5) of that paper fails to be satisfied. Note that
$\delta-r_I\left(1-\frac{p_0}{T_{\rm max}}\right)\ge
\delta\left(1-\max\left\{\frac{r_Id}{r_T\delta},\frac{r_I}{\delta}\right\}
\right)$. Thus sufficient conditions to ensure that the denominator is
positive are that $r_I<\delta$ (maximum proliferation rate of infected cells is
less than their death rate) and $r_T>d$ (proliferation rate of uninfected
cells is greater than their death rate). Alternatively to the second condition
we can assume that $r_T\ge r_I$ (infected cells proliferate slower than
uninfected ones) and $d\le \delta$ (infected cells die faster than uninfected
ones). These conditions
have simple biological interpretations. It is not clear that they must hold in
all biologically interesting situations. It is, for instance, conceivable that
it is of evolutionary advantage for the virus to cause the cells it infects to
proliferate faster than they would otherwise do.

\noindent
{\bf Lemma 3} (i) When $\rho R^*+r_I-\delta\le 0$ the system
(\ref{basic1})-(\ref{basic3}) has no positive steady states.

\noindent
(ii) When $\rho R^*+r_I-\delta>0$ there is a one-to-one correspondence between
positive steady states of the system and roots of a polynomial $p$ in the
interval $(X_-,1)$ where $X_-={\rm max}\left\{0,\hat X\right\}$ and
$\hat X=\frac{c(\delta-r_I)}{b(\rho R^*+r_I-\delta)}$. In the case $X_-\ge 1$
this is to be interpreted as the statement that there are no positive steady
states.

\noindent
{\bf Proof} Consider a steady state $(T^*,I^*,V^*)$ of the system
(\ref{basic1})-(\ref{basic3}) with $R=R^*$. We use the variable $X$ introduced
in the last section. In general this leads to the equations
\begin{eqnarray}
&&s+r_TT^*\left(1-\frac{T^*}{T_{\rm max}X}\right)-dT^*
   -bV^*X=0,\label{Xeq1}\\
&&r_II^*\left(1-\frac{T^*}{T_{\rm max}X}\right)
   +bV^*X-\delta I^*=0,\label{Xeq2}\\
&&\rho R^*I^*-cV^*-bV^*X=0.\label{Xeq3}
\end{eqnarray}
A given steady state defines a value of $X$ in the interval $(0,1)$. Suppose
that conversely $X$ is given and we look for a steady state giving rise to it.
Equation (\ref{Xeq3}) can be rewritten as
\begin{equation}\label{vstar}
V^*=\frac{\rho R^*I^*}{c+bX}.
\end{equation}
Substituting this into (\ref{Xeq2}) and cancelling a factor $I^*$ gives
\begin{equation}
  r_I\left(1-\frac{T^*}{T_{\rm max}X}\right)+\frac{b\rho R^* X}{c+bX}
  -\delta=0.\nonumber
\end{equation}
Hence 
\begin{equation}\label{Tstar}
  T^*=\frac{T_{\rm max}X[b(\rho R^*+r_I-\delta)X+c(r_I-\delta)]}
  {r_I(c+bX)}.
\end{equation}
If $\rho R^*+r_I-\delta$ were not positive then
$r_I-\delta$ would be negative and this relation could not hold for a positive
steady state. In other words $\rho R^*+r_I-\delta>0$ is a necessary condition
for the existence of a positive steady state. This completes the proof of part
(i) of the lemma.

If $r_I\ge \delta$ then the expression (\ref{Tstar}) for $T^*$ is positive.
If, on the other hand $r_I<\delta$ then it is necessary to impose the
restriction $X>\hat X$.
When $T^*$ has been calculated in terms of $X$ it is possible to determine
$I^*=\frac{T^*(1-X)}{X}$ and $V^*$. This shows that there is at most one
steady state consistent with a given value of $X$. The quantities
$(T^*,I^*,V^*)$ defined in terms of $X$ in this way define steady state
solutions of equations (\ref{basic2}) and (\ref{basic3}). Substituting
these equations into (\ref{basic1}) and multiplying the result by
$r_I^2(c+bX)^2$ gives
\begin{eqnarray}
  &&sr_I^2(c+bX)^2+rT_{\rm max}X[b(\rho R^*+r_I-\delta)X+c(r_I-\delta)]
     [-b(\rho R^*+r_I-\delta)X+c\delta]\nonumber\\
  &&+r_IT_{\rm max}X[b(\rho R^*+r_I-\delta)X+c(r_I-\delta)]
     [b (\rho R^*-d)X-dc-b\rho R^*]=0.\nonumber
\end{eqnarray}
Alternatively we can write this equation in the form
\begin{eqnarray}\label{alternative}
  &&sr_I^2(c+bX)^2+T_{\rm max}X[b(\rho R^*+r_I-a)X+c(r_I-\delta)]
     \nonumber\\
  &&\times [b ((\delta r_T-dr_I)-\rho R^*(r_T-r_I))X
     +c(\delta r_T-dr_I)-b\rho R^*r_I)]=0.\label{splitform}
\end{eqnarray}
We can also write it in the form $p(X)=\sum_{i=0}^3b_iX^i=0$ where
\begin{eqnarray}
  &&b_3=T_{\rm max}b^2(\rho R^*+r_I-\delta)[(\delta-\rho R^*)r_T
     -(d-\rho R^*)r_I)],\nonumber\\
  &&b_2=sr_I^2b^2+T_{\rm max}\beta [(\rho R^*+r_I-\delta)(r_Tc\delta
                     -r_I(dc+b\rho R^*))\nonumber\\
  &&+(r_I-\delta)c ((\delta-\rho R^*)r_T-(d-\rho R^*)r_I)],\nonumber
  \\
  &&b_1=2sr_I^2bcT_{\rm max}+(r_I-\delta)c[r_Tc\delta-r_I (dc+b\rho R^*)],\nonumber\\
  &&b_0=sr_I^2c^2.\nonumber
\end{eqnarray}
Suppose now that $X\in (X_-,1)$ satisfies $p(X)=0$. Then it is possible to
define corresponding positive quantities $(T^*,I^*,V^*)$ as above and they
satisfy (\ref{basic2}) and (\ref{basic3}).
Now divide the equation $p(X)=0$ by $r_I^2(c+X)^2$.
Using (\ref{basic2}) and (\ref{basic3}) we can write the expressions depending on $X$ in the
resulting equation in terms of $T^*$ and $V^*$. This shows that (\ref{basic1}) holds.
Thus $(T^*,I^*,V^*)$ is a positive
steady state. It has now been shown that provided $\rho R^*+r_I-\delta>0$
there is a one-to-one correspondence between positive steady states and roots
of the polynomial $p$ in the interval $(X_-,1)$. This is part (ii) of the
lemma. $\blacksquare$

It turns out that a root of the polynomial $p$ can only lie outside the region
where it corresponds to a steady state if ${\cal R}_0''<1$.

\noindent
{\bf Lemma 4} If $\rho R^*+r_I-\delta\le 0$ or $X_-\ge 1$ then ${\cal R}_0''<1$.

\noindent
{\bf Proof}  When $\rho R^*+r_I-\delta\le 0$ it follows that $\rho R^*\le\delta-r_I$ and,
in particular, that $\delta-r_I>0$. It can be concluded that
\begin{equation}
  {\cal R}_0''=\frac{b\rho R^*}{(b+c)\left(\delta-r_I
      +\frac{p_0r_I}{T_{\rm max}}\right)}\nonumber
  \le\frac{b(\delta-r_I)}{(b+c)\left(\delta-r_I
      +\frac{p_0r_I}{T_{\rm max}}\right)}<1.\nonumber
\end{equation} 
When $\rho R^*+r_I-\delta>0$ and $X_-\ge 1$ it follows that
$b\rho R^*\le (b+c)(\delta-r_I)$. Thus
\begin{equation}
  {\cal R}_0''=\frac{b\rho R^*}{(b+c)\left(\delta-r_I
      +\frac{p_0r_I}{T_{\rm max}}\right)}
\le\frac{(b+c)(\delta-r_I)}{(b+c)\left(\delta-r_I
      +\frac{p_0r_I}{T_{\rm max}}\right)}<1.\nonumber
\end{equation}
$\blacksquare$

%The linearization of the system about the disease-free equilibrium was
%computed in \cite{nangue21}. It is given by
%\begin{equation}
%\left[
%  {\begin{array}{ccc}
%     -a_{11} & -a_{12} & -b\\
%     0      & a_{22} & b\\
%     0      & \rho R^* & -b-c\\
%   \end{array}}
%\right]
%\end{equation}
%where $a_{11}=\sqrt{(r_T-d)^2+\frac{4r_Ts}{T_{\rm max}}}$,
%$a_{12}=\frac{p_0r_T}{T_{\rm max}}$ and
%$a_{22}=-\left[\delta+r_I\left(\frac{p_0}{T_{\rm max}}-1\right)\right]$.
%This matrix obviously has an eigenvalue $-a_{11}$. The other two are the roots
%of the polynomial
%$\lambda^2+(-a_{22}+b+c)\lambda-a_{22}(b+c)-b\rho R^*$. The
%product of these two eigenvalues is $-a_{22}(b+c)-b\rho R^*$. It is
%positive when
%\begin{equation}
%(b+c)\left[\delta-r_I\left(1-\frac{p_0}{T_{\rm max}}\right)\right]>b\rho R^*,
%\end{equation}
%which is equivalent to ${\cal R}_0''<1$. The product is negative precisely
%when ${\cal R}_0''>1$. It can be concluded that the steady state has no zero
%eigenvalues except in the case ${\cal R}_0''=1$. We see that $E_0'$ is
%asymptotically stable when ${\cal R}_0''<1$.

It is possible to get some information about the number and stability of
positive steady states. First we compare with the model of \cite{hews21} which
can be obtained from our model by setting $d=s=\eta=0$. In that case for
general values of the parameters there are up to three boundary steady states.
We concentrate on the boundary steady state called $E_i$ in \cite{hews21}.
Its existence corresponds to the fact that $p(X)$ has a factor $X$ when $s=0$.
Its coordinates are
$\left(0,\frac{T_{\rm max}(r_I-\delta)}{r_I},
  \frac{\rho R^* T_{\rm max}(r_I-\delta)}{r_Ic}\right)$.
It evidently exists as a non-negative steady state which is not at the origin
precisely when $r_I>\delta$. In this case $X_-=0$. It is always the case that at
least two of the eigenvalues of the linearization about that point are
negative. The third has the sign opposite to that of
${\cal R}_0''-\frac{r_T}{r_I}$. Thus if
${\cal R}_0''>\frac{r_T}{r_I}$ it is asymptotically stable. Note that in this
case ${\cal R}_0''=\frac{b\rho R^*}{ca}$. Next we perturb the system of
\cite{hews21} by making $s$ slightly positive.

\noindent
{\bf Theorem 7} Suppose that $r_I>\delta$. If ${\cal R}_0''>\frac{r_T}{r_I}$
then for sufficiently small values of $d$ and $s$ with the other parameters
fixed the system (\ref{basic1})-(\ref{basic3}) has a positive steady state
$\hat E_i$ close to
$\left(0,\frac{T_{\rm max}(r_I-\delta)}{r_I},
  \frac{\rho R^* T_{\rm max}(r_I-\delta)}{r_Ic}\right)$. If
${\cal R}_0''<\frac{r_T}{r_I}$ no such steady state exists. An analogous result
holds for the system obtained from (\ref{basic1})-(\ref{basic3}). In that case
$\hat E_i$ is asymptotically stable. 

\noindent
{\bf Proof} By the implicit function theorem the perturbed system has
precisely one steady state close to $E_i$ when $s$ is small enough. It cannot
be on the plane $I=0$ when $s\ne 0$. Consider the case where $E_i$ is
asymptotically stable. Call the perturbed steady state $\hat  E_i$. By
continuity there is an $\epsilon>0$ such that each solution which starts in
$B_\epsilon (\hat E_i)$ converges to $\hat E_i$. When the perturbation is small
enough this ball will intersect the plane $I=0$. Thus there is a solution
which starts on that plane and converges to $\hat E_i$. When the perturbation
and $\epsilon$ are small enough $\dot I$ is positive on that ball.
Hence $\hat E_i$ lies in the positive octant and is asymptotically
stable. If $d$ is increased a little from zero this solution continues to
exist and be asymptotically stable. We can also obtain a corresponding
solution in the case $\eta=1$ by the following trick. We consider the system
obtained from (\ref{basic1})-(\ref{basic3}) by replacing the term
$\frac{bTV}{T+I}$ by $\frac{\eta bTV}{T+I}$. A simple perturbation
argument gives a positive steady state of the system where $\eta$ is a
small positive constant. We can go from there to a steady state with
$\eta=1$ by scaling $\eta$, $\rho$ and $c$ by the same factor, since this
scaling leaves the set of steady states invariant. Note, however, that it is
not clear that this scaling preserves the stability of the steady state.
Note that in the case where $E_i$ is not asymptotically stable there does not
exist a positive steady state $\hat E_i$. To see this consider now the case
where $E_i$ is a saddle. Then there are points on its unstable manifold
arbitrarily close to $E_i$ for which $T$ becomes negative. After a
sufficiently small perturbation this will still be the case. If $\hat E_i$
satisfied the condition $T>0$ then this would give a contradiction since it
would mean that a solution initially in the positive octant leaves the positive
region. $\blacksquare$

In some cases positive steady states can be obtained from steady states on the
boundary by a perturbation analysis, sometimes associated with the term
'backward bifurcation'. The setting is one where the boundary steady state
to be perturbed is such that the reproductive ratio $R_0$ is equal to one. Under
a perturbation of the parameters this point remains a steady state but $R_0$
varies. In some well-known simple cases a positive steady state near the
boundary steady state exists for $R_0>1$ and it is stable. In other cases a
positive steady state exists for $R_0<1$ and is unstable. The latter case has
been called a backward bifurcation. This is because the positive steady state
arises when moving to smaller values of $R_0$ (backwards) rather than (as in
the most familiar case) when moving to larger values of $R_0$ (forwards). An
example where this occurs is an in-host model for HIV studied in \cite{li14}.
There it is proved that there is a backward bifurcation and it is proved
that for $R_0$ slightly less than one there are two positive steady states,
one of which is stable and one unstable. That model includes a description
of therapy appropriate for a reverse transcriptase inhibitor. Mathematically
this means that the coefficient of the infection term in the evolution equation
for $I$ is less than that in the evolution equation for $T$. In that model
mass action kinetics is assumed for the infection and absorption of the
virus is not included.

To investigate which of these phenomena occur in our model we follow the
analysis of \cite{vandendriessche02}. This will be done for the boundary steady
state $E_0'$ where the corresponding reproductive ratio is ${\cal R}_0''$.
We now want to vary a parameter which leaves $E_0'$ invariant but changes
${\cal R}_0''$. This suggests leaving $d$, $r_T$, $s$ and $T_{\rm max}$
unchanged while varying $\delta$, $r_I$, $b$, $\rho$ or $c$. The next step in
the analysis is to find left and right eigenvectors corresponding to the
eigenvalue zero. The vector with components
$(-a_{12}(b+c)-b\rho R^*,a_{11}(b+c),a_{11}\rho R^*)$ is in the
right kernel while that with components $(0,b+c,b)$ is in the left
kernel. The other two eigenvalues are negative. We would now like to compute
the quantities $a$ and $b$ defined in Section 5 of \cite{vandendriessche02}.
As explained there it is enough to know a subset of the second derivatives
of the right hand sides in order to do this. In the terminology of
\cite{vandendriessche02} we choose $I$ and $V$ to be the infected variables
and the term containing $b$ as the only one which contributes to the
matrix $\cal F$. Denote the right hand sides of the evolution equations for
$y$ and $v$ by $f_2$ and $f_3$. Then the relevant second derivatives are
the following derivatives of $f_2$ and $f_3$.
\begin{eqnarray}
  &&\frac{\partial^2 f_2}{\partial I^2}=-\frac{2r_I}{T_{\rm max}}
     +\frac{2b VT}{(I+T)^3},\
  \frac{\partial^2 f_2}{\partial I\partial V}=-\frac{\beta I}{(I+T)^2},\
  \frac{\partial^2 f_2}{\partial V^2}=0,\nonumber\\
  &&\frac{\partial^2 f_3}{\partial I^2}=-\frac{2b VT}{(I+T)^3},\
  \frac{\partial^2 f_3}{\partial I\partial V}=\frac{bT}{(I+T)^2},\
  \frac{\partial^2 f_3}{\partial V^2}=0,\nonumber\\
  &&\frac{\partial^2 f_2}{\partial T\partial I}=-\frac{r_I}{T_{\rm max}}
     +\frac{bV(T-I)}{(I+T)^3},\
  \frac{\partial^2 f_2}{\partial T\partial V}=\frac{bI}{(I+T)^2},\nonumber\\
  &&\frac{\partial^2 f_3}{\partial T\partial I}=-\frac{bV(T-I)}{(I+T)^3},\
  \frac{\partial^2 f_3}{\partial T\partial V}=-\frac{bI}{(I+T)^2}.\nonumber
\end{eqnarray}
These derivatives should be evaluated at the disease-free steady state which
means in particular that it is possible to set $I=0$ and $V=0$, which implies
that $I+T=T$. In the present case the matrix with components $\alpha_{lk}$
introduced in \cite{vandendriessche02} has components
$(0,\frac{\rho R^*}{b+c})$. We now have all the elements necessary to compute
the quantity $a$ introduced in \cite{vandendriessche02} to characterize
backward bifurcations, which we denote by $a_{VW}$.
\begin{eqnarray}
  &&a_{VW}=a_{11}^2(b+c)\left[-\frac{r_I}{T_{\rm max}}(b+c)^2
     -\frac{b\rho R^*c}{I}\right]\nonumber\\
  &&+\frac{a_{11}b(\rho R^*)^2c I}{(b+c)I^2}
     \left[-a_{12}(b+c)-b\rho R^*\right].\nonumber
\end{eqnarray}
This calculation shows that $a_{VW}$ is always positive and thus that no
backward bifurcation occurs in this system. To show that there is a forward
bifurcation it remains only to show that the bifurcation parameter can be
chosen so that the quantity $b$ in \cite{vandendriessche02}, call it $b_{VW}$,
is non-zero. A suitable choice of bifurcation parameter is $\rho-\rho_0$,
where $\rho_0$ is the value of $\rho$ corresponding to ${\cal R}_0''=1$. In that
case $b_{VW}=a_{11}(b+c)\rho R^*>0$. It follows from Theorem 4 of
\cite{vandendriessche02} that for $\gamma$ slightly larger than $\gamma_0$
there is an asymptotically stable positive steady state close to the
bifurcation point. What is more interesting than the fact that a forward
bifurcation occurs here is the fact that no backward bifurcation occurs.

We now study the way in which the number of steady states changes when the
parameters are varied.

\noindent
{\bf Lemma 5} Consider a sequence of positive parameters with
$\rho_nR^*_n+(r_I)_n-\delta_n>0$ for all $n$ converging to a limit
and a sequence $X_n\in ((X_-)_n,1)$ of roots of $p$ for these parameter values
converging to some limit $X^*$. Suppose that $\rho_nR^*_n+(r_I)_n-\delta_n$
does not tend to zero as $n\to\infty$ and that ${\cal R}_0''$ does not tend to
one along the sequence. Then $X^*$ is not equal to the value $X_-^*$ of $X_-$
corresponding to the limiting parameter values or to one.

\noindent
{\bf Proof} Suppose first that $X_n$ tends to $X_-^*$. Then $p(X_-^*)=0$ for the
limiting parameters, contradicting (\ref{alternative}). Next suppose that $X_n$
tends to one. For each $n$ there is a steady state $(T_n,I_n,V_n)$
corresponding to $X_n$ and since $\rho R^*+r_I-\delta$ remains positive in the
limit there is also a corresponding non-negative steady state $(T^*,I^*,V^*)$.
In fact, since $I=\frac{(1-X)T}{X}$, $V^*=I^*=0$ and thus this steady state is
$E_0'$. It follows that along the sequence two steady states approach each
other and that the linearization at the limiting steady state must have zero as
an eigenvalue. This implies that ${\cal R}_0''$ tends to one, contradicting the
assumptions of the lemma. $\blacksquare$

%Note that the analogue of Lemma 4 holds with the same proof when $\rho$ is
%allowed to become zero in the limit.

\noindent
{\bf Theorem 8} If $\rho R^*+r_I-\delta>0$ the number of positive
steady states of the system (\ref{basic31})-(\ref{basic33}) with $\eta=1$
is even for ${\cal R}_0''\le 1$ and odd for
${\cal R}_0''>1$. In particular, there always exists at least one positive
steady state in the latter case.

\noindent
{\bf Proof} Call the given parameter set $z_0$ and suppose that for that
parameter set $\rho=\rho_0$. Consider a family $z(u)$, $u\in [0,1]$, with
$z(0)=z_0$ obtained from $z$ by setting $\rho(u)=(1-u)\rho_0+u\rho_1$ and
leaving the other parameters unchanged. Choose $\rho_1$ so that 
${\cal R}_0''=1$ for $u=1$. Consider first the case ${\cal R}_0''<1$. Then
within the parameter family $\rho$ is an increasing function of $U$ and
$\rho R^*+r_I-\delta$ does not approach zero. It follows from
Lemma 5 that the roots of $p$, which vary continuously with $u$, cannot
approach the endpoints of the interval for $u<1$. Hence the parity of the
number of positive steady states is independent of $\rho$ for ${\cal R}_0''<1$.
Consider next the case ${\cal R}_0''>1$. If $\rho R^*+r_I-\delta$ approached
zero for some $u<1$ then ${\cal R}_0''$ would become less than one as a
consequence of Lemma 4, in
contradiction to the definition of the family. We can then argue as before
to see that the parity of the number of positive steady states is independent
of $\rho$ for ${\cal R}_0''>1$. We have seen what happens when ${\cal R}_0''$
is perturbed a little in the discussion of backward bifurcations. For
${\cal R}_0''$ slightly less than one
the number of roots does not change while for ${\cal R}_0''$ slightly greater
than one it increases by one. Thus the parity changes when ${\cal R}_0''$
passes through one. It remains to determine the parity of the number of
positive steady states for ${\cal R}_0''=1$. In that case $p(X_-)>0$ and 
$p(1)=0$. Thus the parity of the number of roots in the interval of interest
is even. $\blacksquare$
\section{Summary and outlook}\label{outlook}
In \cite{nangue21} an in-host model for hepatitis C was introduced and some
properties of its solutions were determined. At the same time a variety of
questions concerning this model were left open. In the present paper we 
obtain some answers to these questions. In \cite{nangue21} a restriction
on the parameters was identified which implies that every solution converges
to a steady state. It is shown here that convergence to steady states does not
hold without restriction since there exist periodic solutions for some values of
the parameters. Another question is that of the number of steady states.
It is shown that the parity of the number of steady states is determined
by ${\cal R}_0''-1$. This is achieved without any restriction on the parameters
other than their positivity. It follows in particular that for ${\cal R}_0''>1$
there is always at least one positive steady state. It is shown that when this
solution is close enough to the disease-free steady state it is asymptotically
stable.

There remain a number of open questions concerning the model studied in this
paper. Does there ever exist a positive steady state in the case
${\cal R}_0''\le 1$? Does there ever exist more than one positive steady state
in the case ${\cal R}_0''>1$? Despite the fact that a general parametrization
of steady states by the roots of a polynomial has been obtained we did not
succeed in answering these questions. One route which might have led to an
answer is via backward bifurcations. However it turns out that these do not
exist in this model. Related to this is the fact that we do not know if
there ever exist unstable positive steady states in this model. Another route
which might lead to answers to these questions is to investigate the existence
of fold bifurcations in the positive region. At the moment the question of
whether bifurcations of this type occur remains open.

Is is proved that there exist periodic solutions but no proof of their
stability is available. The only indication we have that they are stable
are simulations. It might be possible to investigate their stability 
using the method of Li and Muldowney, generalizing what was done in
\cite{nangue21}. While studying the properties of the model of \cite{nangue21}
some properties of solutions of related models were obtained as by-products.
This concerns in particular the model of Guedj and Neumann \cite{guedj10} but
in that case the question of the existence of periodic solutions was not
settled.

It is clear from this discussion that there remain many open mathematical
questions concerning the model of \cite{nangue21} and related ones. Beyond
this there remain questions concerning the relations of these mathematical
models to the diseases which they are intended to describe. In which phases
of which diseases is which model most appropriate? Once a model has been
chosen in a given biological situation what are appropriate restrictions on
the parameters? We hope that in the future answers to these questions will
lead to a better understanding of hepatitis C and other viral diseases and to
new ideas for treating them.

\vskip 10pt\noindent
{\it Acknowledgments} This work was partially supported by a grant from the
Simons Foundation.

\end{document}